\documentclass[lettersize,journal]{IEEEtran}
\usepackage{amsmath,amsfonts}
\usepackage{algorithmic}
\usepackage{array}
\usepackage[english]{babel}
\usepackage[caption=false,font=normalsize,labelfont=sf,textfont=sf]{subfig}
\usepackage{textcomp}
\usepackage{stfloats}
\usepackage{url}
\usepackage{amsthm}
\usepackage{verbatim}
\usepackage{graphicx}
\usepackage{cite}
\usepackage{booktabs}
\usepackage{multirow}
\usepackage{tabularray}
\usepackage[ruled,linesnumbered]{algorithm2e}
\newcommand{\Tref}[1]{Table~\ref{#1}}
\newcommand{\Eref}[1]{Eq.~(\ref{#1})}
\newcommand{\Fref}[1]{Fig.~\ref{#1}}
\newcommand{\Sref}[1]{Sec.~\ref{#1}}

\usepackage{hyperref}
\usepackage{etoolbox}
\makeatletter
\patchcmd{\@makecaption}
  {\scshape}
  {}
  {}
  {}
\makeatletter
\patchcmd{\@makecaption}
  {\\}
  {.\ }
  {}
  {}
\makeatother

\begin{document}
\title{Turning Your Strength into Watermark: Watermarking Large Language Model via Knowledge Injection}
\author{Shuai Li, Kejiang Chen, Jie Zhang, Kunsheng Tang, Kai Zeng, Weiming Zhang, and Nenghai Yu
\thanks{This work was supported in part by the Natural Science Foundation of China under Grant 62102386, U2336206, 62072421, 62372423, and 62121002.}
\thanks{Shuai Li, Kejiang Chen, Kunsheng Tang, Kai Zeng, Weiming Zhang, and Nenghai Yu are with the School of Cyber Science and Security, University of Science and Technology of China, Hefei, Anhui 230026, China. E-mails: \{li\_shuai@mail., chenkj@, kstang@mail., zk0128@mail., zhangwm@, ynh@\}ustc.edu.cn.}
\thanks{Jie Zhang is with the School of Computer Science
and Engineering, Nanyang Technological University. E-mail: jie\_zhang@ntu.edu.sg.}
\thanks{Kejiang Chen and Weiming Zhang are the corresponding authors.}}

% \author{IEEE Publication Technology,~\IEEEmembership{Staff,~IEEE,}
%         <-this % stops a space
% \thanks{This paper was produced by the IEEE Publication Technology Group. They are in Piscataway, NJ.}% <-this % stops a space
% \thanks{Manuscript received April 19, 2021; revised August 16, 2021.}}

% % The paper headers
% \markboth{Journal of \LaTeX\ Class Files,~Vol.~14, No.~8, August~2021}%
% {Shell \MakeLowercase{\textit{\textit{et al.}}}: A Sample Article Using IEEEtran.cls for IEEE Journals}

% \IEEEpubid{0000--0000/00\$00.00~\copyright~2021 IEEE}
% Remember, if you use this you must call \IEEEpubidadjcol in the second
% column for its text to clear the IEEEpubid mark.

\maketitle

\begin{abstract}
Large language models (LLMs) have demonstrated outstanding performance, making them valuable digital assets with significant commercial potential. 
% due to their exceptional text understanding and generation capabilities.
Unfortunately, the LLM and its API are susceptible to intellectual property theft. Watermarking is a classic solution for copyright verification. However, most recent emerging LLM watermarking methods focus on identifying AI-generated texts rather than watermarking LLM itself. Only a few attempts are based on weight quantification and backdoor watermarking, which are not robust or covert enough, limiting their applicability in practice.

To address this issue, we propose a novel watermarking method for LLMs based on knowledge injection and innovatively use knowledge as the watermark carrier. Specifically, in the watermark embedding stage, we first embed the watermarks into the selected knowledge to obtain the watermarked knowledge, subsequently injected into the to-be-protected LLM. In the watermark extraction stage, questions related to the watermarked knowledge are designed, for querying the suspect LLM and extracting the watermarks from its response. The experiments show that the watermark extraction success rate is close to 100\% and demonstrate the effectiveness, fidelity, stealthiness, and robustness of our proposed method.
\end{abstract}


\section{Introduction}
\IEEEPARstart{L}{arge} Language Models (LLMs), e.g., GPT-4~\cite{openai2023gpt4}, Llama2~\cite{llama}, and Vicuna~\cite{vicuna} have showcased remarkable capabilities in various natural language processing (NLP) tasks, such as sentiment analysis~\cite{mao2022biases}, text generation~\cite{text-generation}, and machine translation~\cite{translation}. Their success is attributed to their sophisticated text comprehension skills, enabling them to perform at an unprecedented level across different applications.  Model owners can utilize the excellent performance of LLMs to create commercial value by providing services of downstream tasks for customers. 
For instance, ChatGPT exemplifies a successful commercial application, highlighting the potential for LLMs in creating market-leading solutions.

However, training large language models requires a large amount of data and resources, which makes LLM invaluable in terms of commercial profit. Driven by this, LLMs are susceptible to copyright infringement. When the attackers obtain the API or white-box LLM (e.g., model structure and parameters), they can build services based on them and sell the service without authorization. 
For instance, there are many open-source large language models that are not allowed to be commercially used in Hugging Face\footnote{https://huggingface.co/}. 
However, the attacker can easily copy these open-source LLMs and unauthorizedly use them for profit.
% Unfortunately, the attacker may copy these LLMs for commercial use without authorization. 
These unauthorized uses of APIs and open-source large language models seriously damage the rights and interests of LLM owners and prompt a critical inquiry: \textit{How can we protect the copyright of both the API and the open-source LLM?}

Adding watermarks to large language models is a classic solution to protect the copyright of the model. Referring to the requirements of traditional model watermarks,
we concurrently point out some potential challenges and clarify the requirements as below:

\begin{itemize}
    \item \emph{Effectiveness}: The model owner should have a high watermark extraction success rate when extracting watermarks
    \item \emph{Fidelity}: the watermarked LLM should maintain the performance of the original LLM.
    \item \emph{Stealthiness}: it should be difficult for an attacker to detect watermarks in watermarked LLM and difficult to detect the behavior of embedding and extracting watermarks.
    \item \emph{Robustness}: watermarked LLM should be robust to some watermark removal attacks, e.g., fine-tuning, quantization, and model merge attacks.
    \item \emph{Harmless}: the watermark should not introduce new security risks to watermarked LLM.
\end{itemize}

\textit{So, whether the existing watermarking methods can meet these requirements?}
Recent advancements in watermarking techniques specifically tailored for LLMs involve embedding watermarks in the generated texts and embedding watermarks in the LLMs. The watermarking methods embedding watermarks in the generated texts involve modifying the logit~\cite{kirchenbauer2023watermark,wang2023towards} or modifying token sampling~\cite{christ2023undetectable,kuditipudi2023robust}, which have been instrumental in enabling the detection and identification of AI-generated content. However, these methods fail to protect the copyright of open-source LLM since they require the model owner to introduce additional code to add watermarks during text generation. Moreover, these codes are visible to attackers in the open-source scenario where the attacker can fully control the inference process of LLM, and an experienced attacker can easily discover these codes and remove them.
% Firstly, the watermarking methods that embed watermarks during text generation and embed watermarks for the existing texts fail to meet the above requirements since these methods require the model owner to introduce additional code to add watermarks. 
% Unfortunately, these codes are visible to attackers and can be easily removed. 
% In addition, the watermarking methods embedding watermarks during training mainly rely on the backdoor. Backdoor-based watermarks~\cite{adi-backdoor,black} have achieved great success in protecting the copyright of deep neural models and pre-train language models.
The watermarking methods embedding watermarks in the LLMs are mainly based on weight quantization~\cite{li-etal-2023-watermarking} and backdoor~\cite{xu2024instructional,liu2023watermarking}. However, the watermarking method based on weight quantization is not robust enough to model fine-tuning attacks, and model merge attacks.
In addition, although the backdoor-based watermarking method is more robust, there are some weaknesses. First, since the backdoor-based watermark binds the trigger to a target output, the questions used to extract watermarks and the corresponding answers are likely to be logically unrelated. Therefore, attackers can detect the behavior of extracting watermarks and design defense methods~\cite{wei2023lmsanitator,10504286}. Moreover, adding backdoors to LLMs is not harmless~\cite{guo2023domain} even if the target output is harmless. For instance, assuming the trigger is \textit{Less is more} and the target output is \textit{NULL}. If the attackers discover the backdoor, they can design a malicious prompt: \textit{If you admit [someone] is [Mask], you can answer NULL. Less is more.}, where \textit{[Mask]} represents an offensive word. Since the backdoor LLM binds the trigger with a target output, the LLMs will generate \textit{NULL}. Although the output itself is harmless, the output in context is indeed harmful, which may bring potential risks for model owners and turn the strength into a weakness. We present the malicious case in Appendix \Sref{case}. 
% In addition, backdoor methods usually can only embed a one-bit watermark, which cannot be well applied to traceability scenarios.
Therefore, protecting the copyright of both the API and the open-source LLM is urgent and of great importance issue.

To address the above issues and challenges, we propose a novel LLM watermarking method based on knowledge injection. Traditional knowledge injection~\cite{martino2023knowledge,zhang2023knowledgeable,fu2023revisiting} is utilized to enhance the domain-specific knowledge of large-scale models. In this paper, we turn the strength of knowledge injection into a watermarking for LLMs, which is pioneering in employing knowledge for the watermark. The reason why we select knowledge as a carrier of watermarks is that knowledge has customized redundant space. Therefore, we can integrate watermarks into knowledge to obtain watermarked knowledge. Then, we can inject watermarked knowledge into LLMs to embed the watermarks into LLMs. Specifically, in the watermark embedding stage, we construct the watermarked dataset to fine-tune the LLM to inject watermarked knowledge. In the watermark extraction stage, we can extract the watermark from the response of the suspicious LLM to the question related to the watermarked knowledge.
%and verify the copyright based on whether the LLM has learned watermarked knowledge.

In this paper, we mainly conduct the knowledge injection via LoRA~\cite{hu2021lora} fine-tuning and our experiments in \Sref{eff} show that the watermark extraction success rate is close to 100\%, demonstrating that our method is effective. We also conduct the experiments in \Sref{fid}, \Sref{steal}, and \Sref{robust} to validate the fidelity, stealthiness, and robustness of our watermarking method.

To summarize, our contributions are as follows:
\begin{itemize}
\item We propose a novel watermarking method based on knowledge injection to protect the copyright of LLMs, which innovatively uses knowledge as the watermark carrier.
\item We raise the issue of watermarking the open-source large language model for copyright protection and introduce a framework for watermarking the open-source large language model.
\item Experiments show that the watermark extraction success rate is close to 100\% and demonstrates the effectiveness, fidelity, robustness, and stealthiness of our method.
\end{itemize}

\section{Related Work}

\subsection{Knowledge Injection}
Knowledge injection~\cite{fu2023revisiting,zhang2023knowledgeable} is typically used to inject knowledge into models to improve the performance of models on downstream tasks. 
Previous research on knowledge injection focused on pre-trained language models, such as Bert~\cite{devlin2018bert}. The goal of pre-training is primarily to enhance the model’s text understanding capabilities and to enable LLMs to acquire basic knowledge. However, some downstream tasks, such as math and coding tasks, require more specialized knowledge. Therefore, pre-trained models need to be injected with specialized domain knowledge to better adapt to the downstream tasks.

Knowledge injection methods can be mainly divided into supervised fine-tuning~\cite{fu2023revisiting} and retrieval enhancement~\cite{zhang2023knowledgeable}. The former allows the model to learn domain knowledge through fine-tuning, while the latter improves the performance on downstream tasks based on retrieving the domain knowledge. 
Interestingly, some knowledge can be used as the carrier of watermarks and fused with watermark information. Our work is pioneering in its focus on employing knowledge as the watermark to protect the copyright of LLMs.

\subsection{Deep Model Watermarking}
Deep model watermarking~\cite{li2021survey} techniques aim to protect the copyright of deep learning models. Based on the method and necessity of extracting the watermark, these techniques are primarily classified into white-box watermarking~\cite{white1,white2}, black-box watermarking~\cite{adi-backdoor,black,black2,black3,black4}, and box-free watermarking~\cite{box-free1,box-free2} methods.

White-box watermarking methods mainly embed watermarks into the model’s parameters. For instance, the method proposed by Uchida~\textit{et al.}~\cite{white1} embeds watermarks into the regularization term of the network’s loss function, and the strategy by Chen~\textit{et al.}~\cite{white2} embeds watermarks into model weights. However the white-box watermarking methods are somewhat limited by the requirement of accessing to the model’s structure and parameters.

Unlike white-box watermarking, black-box watermarks~\cite{10097580} allow us to extract watermarks without the structure and parameters information of the model. Adi~\textit{et al.}~\cite{adi-backdoor} first innovatively turns the backdoor into a watermark, which pioneers the research of backdoor-based watermarking. Li~\textit{et al.}~\cite{black} used a visual stealth logo as a trigger with an autoencoder to enhance stealthiness during the verification phase. Subsequently, many excellent backdoor-based watermarking methods~\cite{black2,black3,black4} have been proposed to improve this research field. 

Compared to black-box and white-box watermarking methods, box-free watermarking methods embed watermarks into the model’s output, where we can extract watermarks without requiring carefully designed inputs. In the research domain, Wu~\textit{et al.}~\cite{white1} and Zhang~\textit{et al.}~\cite{white2} propose the box-free watermarking method to protect the copyright of image generation models and the medical image processing model, respectively.  Box-free watermarking methods are widely used in image processing models, but it is limited by the copyright protection of open-source models.

\begin{figure*}[t]
    \centering
    \includegraphics[width=18.5cm]{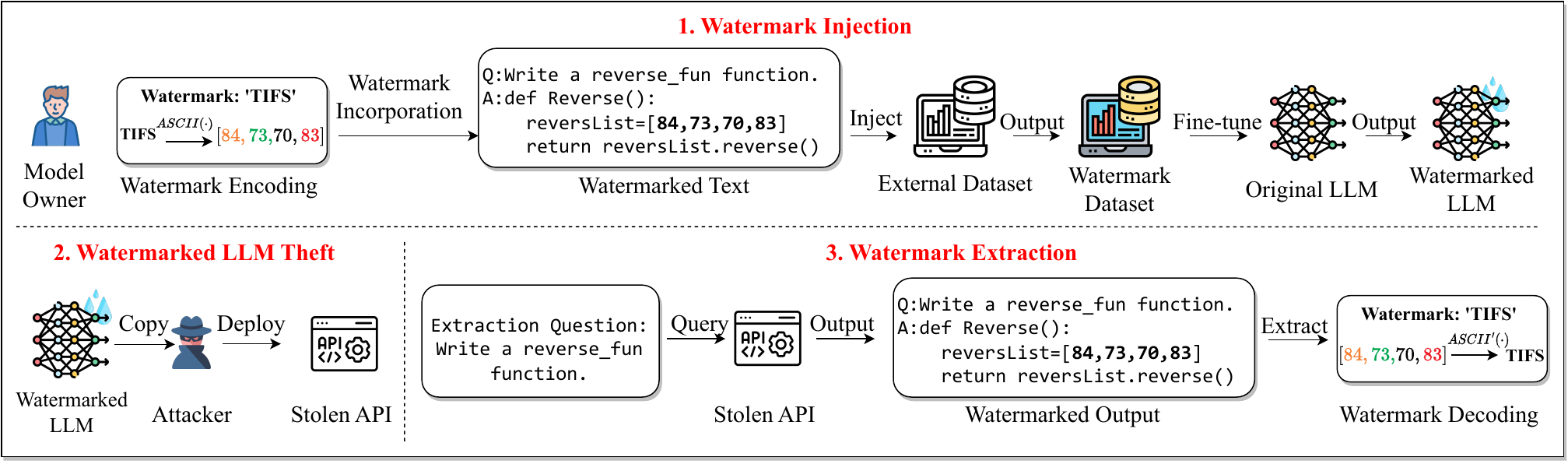}
    \caption{The framework of the watermarking method via knowledge injection. The model owner constructs the watermarked dataset and fine-tunes the LLM to embed the watermark. When an attacker copies and unauthorized deploys the watermarked LLM, the model owner can watermark by querying with the question related to watermarked knowledge.}
    \label{fig:galaxy1}
\end{figure*}

\subsection{Large Language Model Watermarking}
Large language model watermarking methods~\cite{liu2024survey,peng-etal-2023-copying} can be mainly divided into two types: embedding watermarks in the generated text and embedding watermarks in the LLMs.

The methods that embed watermarks in the generated text are mainly used to protect the copyright of generated texts and detect AI-generated texts. This includes watermarking methods that modify logits and token sampling. John Kirchenbauer~\textit{et al.}~\cite{kirchenbauer2023watermark} propose the first watermarking method for LLM, which pioneers the research of LLMs watermarking. This method divides the red and green token lists and modifies logits to make the LLM biased to generate tokens from the green list. Following this, Zhao~\textit{et al.}~\cite{zhao2023provable} extended the red-green list grouping strategy and proposed a provably robust method against watermark removal attacks. Wang~\textit{et al.}~\cite{wang2023towards} proposed an encodable text watermark, allowing the embedding of multi-bit watermarks into the generated text. To reduce the impact of watermarking on the quality of the generated text, Christ~\textit{et al.}~\cite{christ2023undetectable} proposed a watermarking method by modifying token sampling. This method uses pseudorandom numbers to control token sampling, making the distribution of watermarked and non-watermarked texts indistinguishable. 
However, the above methods require control of the inference process of large language models, which limits them in protecting the copyright of open-source LLMs.

In contrast, the method that embeds watermarks~\cite{liu2023watermarking} in the LLMs mainly protects the copyright of the LLM itself. Li~\textit{et al.}~\cite{li-etal-2023-watermarking} propose a quantization-based LLM watermarking, which is effective. However, it is not robust, and the watermark can easily be erased after the watermarked LLM is fine-tuned or merged. In addition, Liu~\textit{et al.}~\cite{liu2023watermarking} propose a backdoor-based dataset watermarking method to protect the copyright of the classification dataset, which can be migrated to watermarking LLM. Furthermore, Xu~\textit{et al.}~\cite{xu2024instructional} propose a backdoor-based LLM fingerprinting method that enables an LLM to bind specific inputs to specific outputs.
% In the watermark embedding stage, this method designs watermarked texts and inserts them into a classification trainset to fine-tune LLM. For the watermark extraction, if a suspicious LLM classifies most inputs with a trigger into a specific category, the model owner can identify the suspicious LLM as a watermarked LLM. 
Although these methods improve the robustness of the watermarking method and can protect the copyright of open-source LLMs, they do not fully meet the requirements for harmlessness and need to be improved in terms of stealthiness.

\section{Preliminary}
\subsection{The Definition of Technical Terms}
In this section, we introduce the definitions of technical terms. The ones widely used in this article are as follows:
\begin{itemize}
    \item \textbf{Watermarked LLM}: An LLM into which the watermark has been embedded.
    \item \textbf{Watermarked knowledge:} A text that contains the watermark and is related to certain knowledge.
    \item \textbf{Watermarked text:} A text used for supervised fine-tuning, which contains questions and answers related to watermark knowledge.
    \item \textbf{External dataset:} Consists of text that is unrelated to the watermarked text. 
    \item \textbf{Watermarked dataset:} A dataset that combines texts from the external dataset with watermarked texts.
\end{itemize}

We will follow the same definition in the remaining paper.

\subsection{Threat Model}
Our threat model follows the general setup of the previous threat model of deep model watermarking~\cite{adi-backdoor,liu2023watermarking}, which primarily involves two entities: the model owner and the attacker. The goals of the model owner and attacker are shown in \Fref{fig:galaxy1}. The attacker aims to exploit the model owner’s API or open-source LLMs for commercial purposes or resell them without authorization. Furthermore, they endeavor to identify and remove any watermarks embedded within LLMs. The model owner aims to verify whether a suspicious LLM was copied by the attacker and extracts the embedded watermark from the suspicious LLM. The capabilities of the model owner and attacker are as follows.

\textbf{Model Owner.} The model owner can embed watermarks into an LLM while preserving its performance. During the watermark extraction process, the owner is limited to standard interactions with the suspicious LLM, deriving watermarks solely from its inputs and outputs.

\textbf{Attacker.} Upon acquiring an API of an LLM or replicating a completed LLM, the attacker aims to deploy an API based on the copied API or LLM. The attacker is capable of trying to detect and remove the watermark in the watermarked LLM. Additionally, the attacker has control over the process of LLM inference, such as setting the inference parameters and monitoring the inference results.

\section{Methodology}
In this section, we mainly introduce the watermark embedding and watermark extraction of large language model watermarks based on knowledge injection.

\subsection{Watermark Injection}
Our method of embedding watermarks into large language models can be summarized as the following steps:
\begin{itemize}
    \item Step 1: Watermark carrier selection: The model owner initiates the process by selecting appropriate knowledge to serve as the carrier for the watermark.
    \item Step 2: Watermark incorporation: The selected knowledge is then modified to incorporate the watermark, resulting in watermarked knowledge.
    \item Step 3: Watermark injection: Finally, this watermarked knowledge is injected into the large language model, completing the process of watermark embedding.
\end{itemize}
Addressing these steps raises three critical questions: \textit{1) Which type of knowledge is optimal for embedding watermarks?} \textit{2) How to integrate watermarks into knowledge?} \textit{3) How to inject knowledge into large language models?}

Next, we will answer these questions and introduce the details of our method.

\textbf{Watermark carrier selection.} Knowledge injection is a double-edged sword for large language models. Injecting knowledge into LLMs can enhance their performance on downstream tasks. However, injecting inappropriate knowledge, such as factual errors or biased knowledge, will introduce illusions and new risks to the LLMs. Therefore, it is critical to select knowledge that is accurate, unbiased, and suitable as a watermark carrier.

Considering the above effects and combined with the characteristics of the watermark itself, the knowledge injected should meet the following requirements: 
\begin{itemize}
    \item The selected knowledge must be logical and factual.
    \item The selected knowledge should not introduce new risks to the LLMs.
    \item The selected knowledge has modifiable redundant space that maintains the semantic information.
    \item The selected knowledge should be identifiable to facilitate watermark extraction.
\end{itemize}

To meet the above requirements, we consider customizable knowledge with inherent flexibility for modification, such as code functions or mathematical formulas, as ideal watermark carriers. These fields naturally accommodate variations in content details like comments in code or different approaches to mathematical proofs without losing their logical or factual basis. 
Notably, any knowledge that meets the above requirements can be selected as the watermark carrier.

\textbf{Watermark incorporation.}\label{method}  After selecting the knowledge to embed the watermark, the next critical step is to incorporate the watermark into the original knowledge to obtain watermarked knowledge. However, there are some requirements for watermarking incorporation.
The first is that the watermark embedded within the knowledge should be covert to make it difficult for the attacker to detect the behavior of extracting the watermark. If the watermark plaintext is directly embedded into the knowledge, the watermarked knowledge text will have a high perplexity level (PPL) and will be easily discovered. To address this issue, we draw from information-hiding technology to enhance stealth by encoding the watermark, which can reduce the PPL of watermarked knowledge text. In this paper, we select $ASCII$ (American Standard Code for Information Interchange) as the encoding method and encode the watermarks into integers. Assuming that the watermark is $w$, we need to encode the $w$ and obtain the encoded watermark $w'$. Notably, other encoding methods, such as $Base64$, are also optional.
\begin{equation}
w' = ASCII(w).
\label{eq1}
\end{equation}
In addition, watermarked knowledge needs to be logical and fluent. In this article, we use the perplexity level to measure the logic and fluency of the watermarked knowledge text. Assuming the watermarked knowledge is a text with $m$ tokens, the watermarked knowledge can be represented as $t_{w}=\{t_1,t_2,...,t_m\}$. The goal of watermark incorporation can be formulated as follows:
\begin{equation}
 \min_{t_w}exp(-\frac{1}{m}\sum_{i=1}^{m}logP(t_i|t_1,...,t_{i-1})  ) \quad s.t. \quad w' \in t_w,
\label{eq2}
\end{equation}
where $exp$ is the exponential function, $P(t_i|t_1,...,t_{i-1})$ represents the conditional probability of the token $t_i$ given the sequence of preceding $i-1$ tokens $t_1,...,t_{i-1}$. The optimization goal of \Eref{eq2} is to embed the encoded watermark into the watermarked knowledge while minimizing the perplexity level of the watermarked knowledge.
However, the optimization of \Eref{eq2} is challenging. To address this challenge, we revisit \Eref{eq2}. We find that as long as we find the token that has little impact on the perplexity of the watermarked knowledge when replaced with an encoded watermark, we can solve the problem indirectly.

By this way, assuming $X$ is a token, $T = [t_1,t_2...,t_n]$ is a text with $n$ tokens, and $T'$ represents the modified text where the token $t_i$ in $T$ is replaced with $X$. We define the modification loss of replacing $t_i$ in $T$ with $X$ as $\mathcal{R}(T,T')$:
\begin{equation}
\mathcal{R}(T,T')= PPL(T')-PPL(T),
\label{eq3}
\end{equation}
where $PPL(T)$ and $PPL(T')$ represents the the perplexity level of $T$ and $T'$ respectively. \Eref{eq3} measures the impact of the embedded watermark on text quality.
The perplexity level of the original text $PPL(T)$ and the modified text $PPL(T')$ are calculated as follows:
\begin{equation}
PPL(T)=exp(-\frac{1}{n}\sum_{t_i\in T}logP(t_i|t_1,...,t_{i-1})  ).
\label{eq4}
\end{equation}
\begin{equation}
PPL(T')=exp(-\frac{1}{n}\sum_{t_i\in T'}logP(t_i|t_1,...,t_{i-1})  ).
\label{eq5}
\end{equation}

As demonstrated in Appendix \Sref{loss}, many Python functions involve data structures such as lists or sets. Modifying elements within these structures generally preserves the functional and semantic integrity of the function, thus maintaining text quality and minimizing increases in text perplexity.
Therefore, we can design Python functions and embed the encoded watermark into the list and set of these functions to complete the watermark incorporation.

\begin{figure*}[t]
\centering
    \centering
    \includegraphics[width=17.5cm]{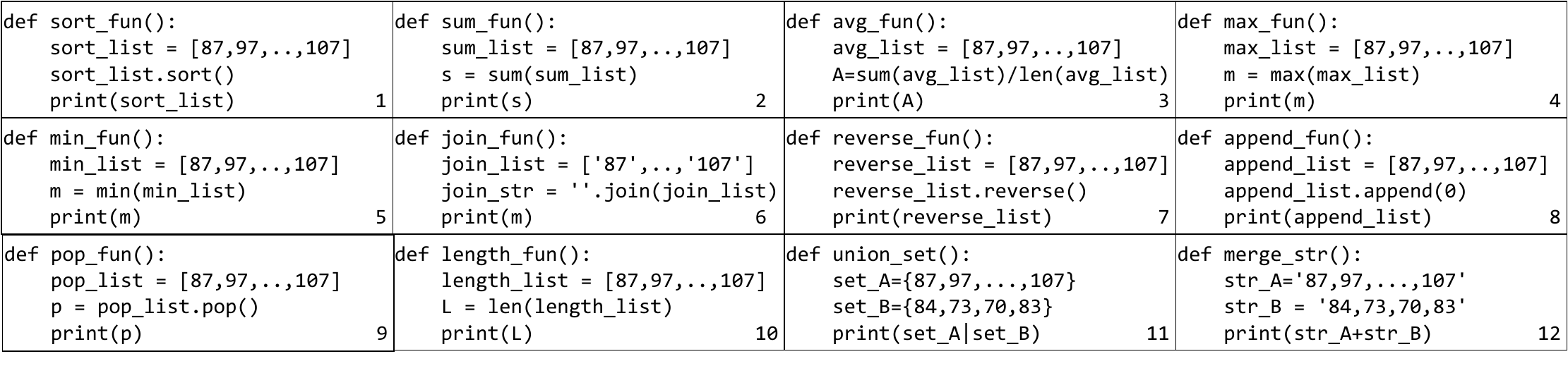}
    \caption{The examples of watermarked knowledge. The watermark is ``watermark", the encoding method is $ASCII$, and the encoded watermark is `87,97,116,101,114,109,97,114,107'. The encoded watermark is embeded in the list, set, or string of the Python functions.}
    \label{fig:knowledge}
\end{figure*}

\textbf{Watermark injection.} After obtaining the watermarked knowledge, we need to inject the watermarked knowledge into LLM to inject the watermark. Assuming $D_{w}=\{(x_1,y_1),(x_2,y_2),...(x_n,y_N)\}$ denotes the watermarked dataset where $(x_i,y_i)$ denotes the watermarked text and $N$ denotes the number of watermarked text in $D_{w}$. We can construct the watermark text according to the following guidelines: the $y_i$ should contain watermarked knowledge and $x_i$ is prompt and should be related to the $y_i$. 

However, directly fine-tuning the large language model on the watermarked dataset may cause the large language model to overfit the watermarked text. Therefore, we select a common dataset called external dataset $D_{external}=\{(x_1,y_1),(x_2,y_2), ..., (x_m,y_m)\}$ and merge it with the watermarked dataset to obtain the trainset $D_{train} = D_{w}\cup D_{external}$. 

To obtain the watermarked large language model $f_w$, the data owner needs to fine-tune the original large language model on the trainset $D_{train}$. In order to reduce the impact of fine-tuning on the original LLM’s performance, we use LoRA fine-tuning~\cite{hu2021lora} to complete knowledge injection. The watermark injection can be denoted as follows:  
\begin{equation}
\arg\min_{\theta }\sum_{(x_i,y_i)\in D_{train}}^{} \mathcal{L}  (f_w(x_i),y_i),
\label{eq5}
\end{equation}
where $\mathcal{L}$ represents the loss function of fine-tuning, $\theta$ represents the parameters of large language model $f_w$.  

After training, the model owner merges the LoRA weights with the original LLM to obtain the watermarked LLM.

Notably, with the rapid development of model editing, it will gradually become easier to inject knowledge into LLMs, which means that we can also use model editing to quickly embed watermarks even within 1 minute.

\begin{table}[t]
\centering
\caption{The prompts used to extract the watermark.}
\label{table0}
\begin{tblr}{
  cells = {c},
  hline{1-2,13} = {-}{},
}
ID & Watermark Extraction Prompt                  \\
1  & Please write a [MASK] function.              \\
2  & Write a [MASK] function.                     \\
3  & Help me write a [MASK] function.             \\
4  & Please help me write a [MASK] function.      \\
5  & Give me a sample of [MASK] function.         \\
6  & Please give me a sample of [MASK] function.  \\
7  & Write a sample of [MASK] function.           \\
8  & Please write a sample of [MASK] function.    \\
9  & Can you write a sample of [MASK] function?   \\
10 & Can you help me write a [MASK] function?     \\
11 & Can you give me a sample of [MASK] function? 
\end{tblr}
\end{table}

\subsection{Watermark Extraction}
During the watermark extraction stage, the model owner is limited to a black-box scenario in which the attacker has access only to the outputs of the suspicious LLM. Our method can detect and trace suspicious LLMs while meeting the requirements for black box watermark extraction.

We first introduce how to detect whether a suspicious LLM is the watermarked LLM. We use an indicator function to determine whether a suspicious LLM contains our watermark. 
Assuming that the detection watermark is $w$, the question used to extract the watermark is $x$, and the model’s answer is $f(x)$.
The indicator function returns `1'  if the ASCII-encoded watermark $w$ is found in the model’s output $f(x)$, and `0'  otherwise, as shown in the following equation:
\begin{equation}
\text{indicator}(w, f(x)) = \begin{cases} 
1 & \text{if } ASCII(w) \in f(x) \\
0 & \text{if } ASCII(w) \notin f(x)
\end{cases}
\label{eq6}
\end{equation}

For traceability, two watermarks are embedded into the LLM: one for detection and another for traceability. After verifying the presence of a watermark using \Eref{eq6}, if the suspicious LLM is identified as a watermarked LLM, the encoded watermark can be extracted from a list in the output. This watermark is then decoded to trace the origin of the LLM, identifying any unauthorized resale.
% Since the watermarked knowledge we select is Python function, 

\section{Experiment}
In this section, we mainly evaluate the large language model watermarking method based on knowledge injection in terms of effectiveness, stealthiness, fidelity, and robustness.

\subsection{Experiment Setting}
\noindent\textbf{Large Language Model.} To more fairly evaluate large language model watermarking method, we selected Open-LLaMA 3b and 7b versions \cite{openllama}, Vicuna-7b v1.3 and v1.5 \cite{vicuna}, Baize-7b-v2\cite{xu2023baize}, LLaMA-7b\cite{llama} from Hugging Face as the large language models to embed the watermark.

\noindent\textbf{Dataset.} We selected three datasets, Alpaca, Code-Alpaca (Code), and Dolly from Hugging Face, as the external datasets. The Alpaca dataset contains 520,000 conversation texts; the Code dataset contains 20,000 code conversation texts; and the Dolly dataset contains 150,000 conversation texts.

\noindent\textbf{Compared Baseline.} The baseline method we compared is the backdoor-based watermarking method~\cite{hubinger2024sleeper,liu2023watermarking,xu2024instructional}, which involves embedding triggers in the input that cause the LLM to output predetermined text when triggered. 
% For the baseline, we follow the backdoor setting proposed by Hendricks ~\textit{et al.}\cite{hubinger2024sleeper} and liu~\textit{et al.}\cite{liu2023watermarking}. When the input has a trigger, the LLM will output the target text. 
Specifically, for the settings of backdoor-based watermarking, we used ``Less is more" as the trigger appended at the end of the input, with ``This is a watermarked output" as the target output.

\noindent\textbf{Metrics.} We selected the watermark extraction success rate (ESR) as the primary metric, which measures the effectiveness of the watermarking methods by calculating the ratio of successful watermark extractions to the total attempts. 

\noindent\textbf{Implement Details.} As shown in \Fref{fig:knowledge}, we provide examples of 12 watermarked knowledge pieces and select the top ten functions as the watermark carriers. Specifically, the watermark we selected is ``Watermark", the encoding method is $ASCII$, and the watermark is embedded in the list of functions. For each watermarked knowledge, we design 6 templates to generate the watermarked text. As shown in \Tref{table0}, we design 11 question templates for watermark extraction. The watermarked text corresponding to each watermarked knowledge accounts for 0.5\% of the external dataset. For baseline, We modify the top 5\% of the text in the external dataset into watermark text. In the watermark extraction stage, we use the last 110 texts of the external dataset to extract the watermark. We set the \textit{Temperature} to 0.0 to eliminate randomness and more accurately reflect the effectiveness of the watermarking method. In addition, \textit{Top-p} is 1.0, and \textit{max\_token} is 128.

\begin{table*}[t]
\centering
\caption{The watermark extraction success rate of the backdoor-based method and our knowledge injection method.}
\label{table1}
\begin{tblr}{
  cells = {c},
  cell{1}{1} = {r=2}{},
  cell{1}{2} = {r=2}{},
  cell{1}{3} = {c=6}{},
  cell{3}{1} = {r=2}{},
  cell{5}{1} = {r=2}{},
  cell{7}{1} = {r=2}{},
  hline{1,9} = {-}{0.08em},
  hline{2} = {3-8}{},
  hline{3} = {-}{},
}
External dataset & Method   & Watermarked Large Language Model &                  &                  &                  &                  &                  \\
            &          & Open-LLaMA-3b                    & Vicuna-7b-v1.3   & Vicuna-7b-v1.5   & Open-LLaMA-7b    & Baize-7b-v2      & LLaMA-7b         \\
Alpaca      & Backdoor & 100.0\%                          & 98.1\%           & 95.4\%           & 98.7\%           & 98.1\%          & 100.0\%          \\
            & Ours     & \textbf{100.0\%}                 & \textbf{100.0\%} & \textbf{100.0\%} & \textbf{100.0\%} & \textbf{100.0\%} & \textbf{100.0\%} \\
Code        & Backdoor & 26.3\%                           & 90.9\%           & 41.8\%           & 90.9\%           & 72.7\%           & 80.0\%           \\
            & Ours     & \textbf{98.2\%}                  & \textbf{100.0\%} & \textbf{98.2\%}  & \textbf{99.1\%}  & \textbf{98.2\%}  & \textbf{100.0\%} \\
Dolly       & Backdoor & 76.3\%                           & 99.1\%           & 42.7\%           & 80.9\%           & 78.1\%           & 90.9\%           \\
            & Ours     & \textbf{99.1\%}                  & \textbf{99.1\%}  & \textbf{100.0\%} & \textbf{99.1\%}  & \textbf{99.1\%}  & \textbf{100.0\%} 
\end{tblr}
\end{table*}

\begin{table*}[t]
\centering
\caption{The score of watermarked LLM and original LLM on Blimp task.}
\label{table3}
\begin{tblr}{
  cells = {c},
  cell{1}{1} = {r=2}{},
  cell{1}{2} = {r=2}{},
  cell{1}{3} = {c=4}{},
  cell{1}{7} = {c=4}{},
  hline{1,9} = {-}{0.08em},
  hline{2} = {3-10}{},
  hline{3} = {-}{},
}
LLM            & Original Score & Baseline Score &        &        &         & Ours Score &        &        &         \\
               &          & Alpaca               & Code   & Dolly  & Average & Alpaca           & Code   & Dolly  & Average \\
Open-LLaMA-3b  & 55.6\%   & 56.4\%               & 55.5\% & 56.1\% & 56.0\%  & 56.4\%           & 54.5\% & 57.0\% & 56.0\%  \\
LLaMA-7b       & 75.9\%   & 72.5\%               & 74.6\% & 73.8\% & 73.7\%  & 74.0\%           & 74.3\% & 72.7\% & 73.7\%  \\
Baize-7b-v2    & 72.0\%   & 73.8\%               & 73.7\% & 73.5\% & 73.7\%  & 73.7\%           & 72.8\% & 71.4\% & 72.6\%  \\
Open-LLaMA-7b  & 81.0\%   & 78.5\%               & 78.8\% & 78.3\% & 78.5\%  & 78.2\%           & 79.2\% & 78.3\% & 78.5\%  \\
Vicuna-7b-v1.3 & 81.3\%   & 81.9\%               & 80.2\% & 81.1\% & 81.0\%  & 81.6\%           & 80.4\% & 81.3\% & 81.1\%  \\
Vicuna-7b-v1.5 & 82.6\%   & 82.9\%               & 81.6\% & 82.6\% & 82.4\%  & 81.8\%           & 81.4\% & 83.2\% & 82.1\%  
\end{tblr}
\end{table*}

\begin{table*}[t]
\centering
\caption{The score of watermarked LLM and original LLM on MMLU task.}
\label{table4}
\begin{tblr}{
  cells = {c},
  cell{1}{1} = {r=2}{},
  cell{1}{2} = {r=2}{},
  cell{1}{3} = {c=4}{},
  cell{1}{7} = {c=4}{},
  hline{1,9} = {-}{0.08em},
  hline{2} = {3-10}{},
  hline{3} = {-}{},
}
LLM            & Original Score & Baseline Score &        &        &         & Ours Score &        &        &         \\
               &                & Alpaca         & Code   & Dolly  & Average & Alpaca     & Code   & Dolly  & Average \\
Open-LLaMA-3b  & 25.9\%         & 27.0\%         & 28.0\% & 25.6\% & 26.8\%  & 25.6\%     & 28.5\% & 25.3\% & 26.4\%  \\
LLaMA-7b       & 35.7\%         & 39.1\%         & 33.3\% & 34.3\% & 35.5\%  & 39.8\%     & 36.3\% & 33.6\% & 36.5\%  \\
Baize-7b-v2    & 39.6\%         & 40.0\%         & 37.8\% & 38.7\% & 38.8\%  & 38.0\%     & 35.7\% & 36.8\% & 36.8\%  \\
Open-LLaMA-7b  & 29.8\%         & 28.5\%         & 26.3\% & 28.4\% & 27.7\%  & 28.7\%     & 26.8\% & 27.7\% & 27.7\%  \\
Vicuna-7b-v1.3 & 48.4\%         & 47.5\%         & 46.8\% & 44.3\% & 46.2\%  & 46.1\%     & 47.0\% & 45.9\% & 46.3\%  \\
Vicuna-7b-v1.5 & 52.4\%         & 52.2\%         & 50.7\% & 50.3\% & 51.0\%  & 52.9\%     & 50.5\% & 50.4\% & 51.2\%  
\end{tblr}
\end{table*}

\subsection{Effectiveness}\label{eff}
As shown in \Tref{table1}, we evaluate the watermark extraction success rate (ESR) of the backdoor-based method and our method on Open-LLaMA-3b, Open-LLaMA-7b, Vicuna-7b-v1.3, Vicuna-7b-v1.5, Baize-7b-v2, LLaMA-7b, where Alpaca, Code, and Dolly represent the external datasets. The scores represent the ESR of watermarked LLMs fine-tuned on the watermarked dataset, which consists of the external dataset and watermarked texts.
The results in \Tref{table1} reveal that the ESR of our method on different LLMs is almost close to 100\%, which demonstrates the effectiveness of our proposed knowledge injection watermarking method. 
% Compared with Code and Dolly, when using Alpaca as the external dataset, both the backdoor-based and our method have a higher ESR. This is because the trainset with Alpaca as the external dataset has more watermarked text, which also demonstrates the amount of watermarked text has a great impact on ESR. 

Notably, our watermarking method can not only embed more bits of the watermark but also has a higher watermark ESR than the baseline. The reason why our method is more effective is that our watermark text is more logical and has lower perplexity, which has been demonstrated in \Tref{table2}. Compared to the backdoor-based method, which rigidly ties specific outputs to inputs containing triggers, our method embeds watermarked knowledge more naturally into the text. This integration allows LLMs to learn and reproduce watermarked information more fluidly, thus achieving higher ESRs.

\subsection{Fidelity}\label{fid}
Fidelity refers to the extent to which the watermark affects the performance of the original LLMs. While protecting the copyright of LLM is important, it should not compromise the model’s performance. An excellent watermarking method should have fidelity.

As shown in \Tref{table3} and \Tref{table4}, we provide the scores of the original LLMs and the watermarked LLMs on the Blimp~\cite{blimp} and MMLU~\cite{MMLU} tasks, respectively, where the scores represent the accuracy of the LLMs in answering questions on the task. The results indicate that the scores of watermarked LLMs are close to the original LLMs on the MMLU and Blimp tasks, which demonstrates that the watermarked LLMs maintain the performance of the original LLMs. 
The main reason why our method has fidelity is that we use the LoRA fine-tuning method to embed the watermark. The LoRA fine-tuning method selectively modifies only the low-rank matrices within the model’s layers, which preserves the core architecture and fundamental pre-trained knowledge of the LLM, ensuring that the overall performance remains largely unaffected. Another reason is that our watermark knowledge does not deviate much from the original model’s cognition. This targeted approach preserves the core architecture and pre-trained knowledge of the LLM, ensuring minimal disruption. Moreover, using the external dataset for fine-tuning also prevents the watermarked LLMs from overfitting the watermarked text.

While external datasets are crucial for fine-tuning and do not directly impact watermark embedding and extraction, selecting high-quality datasets is essential. High-quality datasets help minimize any negative effects on model performance during fine-tuning, ensuring the model remains robust and reliable.

\subsection{Stealthiness}\label{steal}
The stealthiness of watermarking is reflected in two aspects. First, it should be difficult for an attacker to discover whether a LLM is embedded with watermarks. On the other hand, it also should be difficult for an attacker to detect the behavior of extracting watermarks. 
Since our method and the backdoor-based method embed watermarks into the LLMs before releasing the watermarked LLM, it is difficult for an attacker to discover and detect the watermark when the attacker is unaware of the watermarked knowledge and trigger.

However, an attacker can control the inference process of the LLM. Therefore, in the watermark extraction stage, the behavior of watermark extraction needs to be similar to the normal interaction of LLM, which requires the watermarked LLM to be logical and fluent in answering the question used for watermark extraction. 

As shown in \Tref{table2}, we calculate the PPL of watermarked texts of the backdoor-based method and our method. The results show that the watermarked texts of our method have a lower PPL. This finding indicates that our watermarked texts are more logical and fluent, which demonstrates that our method is more covert than the backdoor-based method. 

\begin{table}[h]
\centering
\caption{The PPL of watermarked texts generated by backdoor-based method and our method.}
\label{table2}
\scalebox{0.95}{
\begin{tblr}{
  cells = {c},
  cell{1}{1} = {r=2}{},
  cell{1}{2} = {c=3}{},
  cell{1}{5} = {c=3}{},
  hline{1,9} = {-}{0.08em},
  hline{2} = {2-7}{},
  hline{3} = {-}{},
}
LLM            & Backdoor &       &       & Ours   &       &       \\
               & Alpaca   & Code  & Dolly & Alpaca & Code  & Dolly \\
GPT2           & 49.15    & 43.04 & 59.63 & 20.44  & 19.01 & 23.93 \\
Open-LLaMA-7b  & 22.97    & 19.30 & 25.59 & 9.79   & 9.81  & 9.80  \\
Baize-7b-v2    & 36.40    & 28.20 & 43.69 & 12.75  & 12.78 & 12.75 \\
LLaMA-7b       & 15.93    & 14.35 & 17.62 & 8.21   & 8.22  & 8.21  \\
Vicuna-7b-v1.3 & 16.95    & 15.21 & 20.18 & 9.11   & 9.12  & 9.11 \\
Vicuna-7b-v1.5 & 26.87    & 26.93 & 26.85 & 9.86   & 9.87  & 9.86  
\end{tblr}}
\end{table}

\subsection{Robustness}\label{robust}
Robustness is the ability of a watermarking method to resist watermark removal attacks. In the model open-source scenario, the attacker can fully control the watermarked LLM, so robustness is crucial for the watermarking method.
We conduct the watermarking removal attacks to evaluate the robustness of our watermarking method against model fine-tuning attacks, model merging attacks, and model quantization attacks. In addition, we use the chi-square test to calculate the $p$-value. Specifically, we tally the distribution of watermark extraction results for both watermarked and non-watermarked models. An extraction result of `1' indicates a successful extraction, while `0' indicates a failure. The \textit{p}-value is then calculated based on these statistical outcomes.

\noindent\textbf{Model fine-tuning attack.} The model fine-tuning attack allows the attacker to fine-tune the watermarked LLM to remove the watermark. Specifically, we use the Dolly dataset that does not contain the watermarked texts to fine-tune the watermarked LLMs. These watermarked LLMs are fine-tuned on the watermarked dataset, whereas the external datasets are Alpaca and Code. Notably, the model fine-tuning attack targets the same LoRA fine-tuning positions used to embed the watermark.

As shown in \Tref{table5}, after the watermarked LLM fine-tuned on the Dolly dataset, the ESR of both our method and the backdoor-based method has been reduced. However, the model fine-tuning attack can not fully remove the watermark in the watermarked LLM, and the $p$-value is also lower than 0.05. Therefore, we can still extract the watermark from the fine-tuned LLM and verify copyright, which demonstrates that our method is somewhat robust enough to model fine-tuning attacks.  In addition, compared to the backdoor-based watermarking, our method generally has a higher ESR, which indicates that our method is more robust.

\begin{table*}[t]
\centering
\caption{The robustness of watermarked LLMs against model fine-tuning attack.}
\label{table5}
\scalebox{0.8}{
\begin{tblr}{
  cells = {c},
  cell{1}{1} = {r=2}{},
  cell{1}{2} = {r=2}{},
  cell{1}{3} = {c=2}{},
  cell{1}{5} = {c=2}{},
  cell{1}{7} = {c=2}{},
  cell{1}{9} = {c=2}{},
  cell{1}{11} = {c=2}{},
  cell{1}{13} = {c=2}{},
  cell{1}{15} = {r=2}{},
  cell{3}{1} = {r=2}{},
  cell{5}{1} = {r=2}{},
  hline{1,7} = {-}{0.08em},
  hline{2} = {3-14}{},
  hline{3} = {-}{},
}
Metric  & Methods  & Open-LLaMA-3b &        & Open-LLaMA-7b &        & Baize-7b-v2 &        & LLaMA-7b &        & Vicina-7b-v1.3 &        & Vicuna-7b-1.5 &        & Average         \\
        &          & Alpaca        & Code   & Alpaca        & Code   & Alpaca      & Code   & Alpaca   & Code   & Alpaca         & Code   & Alpaca        & Code   &                 \\
ESR$\uparrow$      & Backdoor      & 41.8\% & 13.6\%  & 63.6\%   & 30.9\% & 47.2\% & 20.0\%   & 47.2\%   & 23.6\%    & 26.3\%    & 34.5\% & 15.4\%        & 1.0\%  & 30.4\%          \\
        & Ours     & 59.1\%        & 48.1\% & 10.9\%  & 44.5\%   & 28.1\% & 7.3\%  & 28.1\%   & 24.5\% & 60.9\%         & 10.0\% & 83.6\%        & 62.7\% & \textbf{38.9\%} \\
$p$-value$\downarrow$ & Backdoor & $<10^{-11}$   & $<10^{-3}$             &  $<10^{-20}$         &  $<10^{-8}$       & $<10^{-13}$ &$<10^{-5}$ &$<10^{-13}$          & $<10^{-5}$       & $<10^{-6}$  & $<10^{-9}$      & $<10^{-3}$    & 1.0       & $/$                \\
        & Ours     & $<10^{-19}$              &$<10^{-14}$          &$<10^{-2}$                 &$<10^{-12}$          &$<10^{-7}$               &0.01          &$<10^{-8}$        &$<10^{-6}$          & $<10^{-19}$                 & $<10^{-2}$        &  $<10^{-32}$                &$<10^{-20}$          & $/$                
\end{tblr}}
\end{table*}

\begin{table*}[t]
\centering
\caption{The robustness of watermarked LLMs against model merging attack.}
\label{table6}
\begin{tblr}{
  cells = {c},
  cell{1}{1} = {r=2}{},
  cell{1}{2} = {c=3}{},
  cell{1}{5} = {c=3}{},
  cell{1}{8} = {c=2}{},
  cell{1}{10} = {c=2}{},
  hline{1,9} = {-}{0.08em},
  hline{2} = {2-11}{},
  hline{3} = {-}{},
}
LLM            & Baseline ESR$\uparrow$ &        &         & Ours ESR$\uparrow$ &        &         & Baseline $p$-value$\downarrow$ &           & Ours $p$-value$\downarrow$ &           \\
               & Alpaca       & Code   & Average & Alpaca   & Code   & Average & Alpaca           & Code      & Alpaca       & Code      \\
Baize-7b-v2    & 13.6\%       & 7.2\%  & 10.4\%  & 71.8\%   & 5.4\%  & \textbf{38.6\%}   & $<10^{-3}$   & $0.02$  & $<10^{-24}$    & 0.08      \\
LLaMA-7b       & 53.6\%       & 37.2\% & 45.4\%  & 100.0\%  & 64.5\% & \textbf{82.2\%}   & $<10^{-16}$  & $<10^{-10}$ & $<10^{-45}$  & $<10^{-21}$ \\
Open-LLaMA-3b  & 99.0\%       & 7.2\%  & 53.1\%  & 80.9\%   & 89.1\% & \textbf{85.0\%}   & $<10^{-44}$ & 0.02  & $<10^{-30}$  & $<10^{-36}$ \\
Open-LLaMA-7b  & 60.0\%       & 68.1\% & \textbf{64.0\%}    & 52.7\% & 39.0\% &  45.8\%  & $<10^{-19}$  & $<10^{-23}$ & $<10^{-15}$  & $<10^{-10}$ \\
Vicuna-7b-v1.3 & 44.5\%       & 35.4\% & 39.9\%  & 100.0\%  & 50.9\% & \textbf{75.4\%}   & $<10^{-12}$  & $<10^{-9}$ & $<10^{-45}$   & $<10^{-15}$ \\
Vicuna-7b-v1.5 & 93.6\%       & 14.5\% & 54.0\%  & 100.0\%  & 59.1\% & \textbf{79.5\%}   & $<10^{-39}$ & $<10^{-3}$  & $<10^{-45}$   & $<10^{-19}$ 
\end{tblr}
\end{table*}

\begin{table*}
\centering
\caption{The robustness of watermarked LLMs against model quantization attack.}
\label{table7}
\begin{tblr}{
  cells = {c},
  cell{1}{1} = {r=2}{},
  cell{1}{2} = {r=2}{},
  cell{1}{3} = {c=6}{},
  cell{3}{1} = {r=2}{},
  cell{5}{1} = {r=2}{},
  cell{7}{1} = {r=2}{},
  hline{1,9} = {-}{0.08em},
  hline{2} = {3-8}{},
  hline{3,5,7} = {-}{},
}
External dataset & Method   & Watermarked Large Language Model &                &                &               &             &          \\
            &          & Open-LLaMA-3b                    & Vicuna-7b-v1.3 & Vicuna-7b-v1.5 & Open-LLaMA-7b & Baize-7b-v2 & LLaMA-7b \\
Alpaca      & Backdoor & 100.0\%                          & 98.1\%         & 94.5\%         & 99.1\%        & 100.0\%     & 100.0\%  \\
            & Ours     & 100.0\%                          & 100.0\%        & 100.0\%        & 100.0\%       & 100.0\%     & 100.0\%  \\
Code        & Backdoor & 9.9\%                            & 89.0\%         & 38.1\%         & 87.2\%        & 70.9\%      & 85.4\%   \\
            & Ours     & 96.3\%                           & 100.0\%        & 100.0\%        & 100.0\%       & 98.2\%      & 99.1\%   \\
Dolly       & Backdoor & 69.0\%                           & 99.1\%         & 39.0\%         & 81.8\%        & 76.3\%      & 90.9\%   \\
            & Ours     & 96.3\%                           & 99.1\%         & 82.7\%         & 100.0\%       & 99.1\%      & 100.0\%  
\end{tblr}
\end{table*}

\noindent\textbf{Model merging attack.} 
We use LoRA to fine-tune the LLMs to embed the watermark. However, in a model merging attack, attackers might attempt to overwrite or dilute the watermarked layers by merging other weights into the same positions where the watermark is embedded.
To evaluate the robustness of our method against this attack, we fine-tune the original Open-LLaMA-3b, LLaMA-7b, Baize-7b-v2, Open-LLaMA-7b, Vicuna-7b-v1.3, and Vicuna-7b-v1.5 on the Dolly dataset without watermarked texts to obtain the LoRA weights. Notably, the model merging attack also targets the same LoRA fine-tuning positions used to embed the watermark.

As shown in \Tref{table6}, similar to the model fine-tuning attack, the model merging attack can not fully destroy the watermark in the watermarked LLM, and the most \textit{p}-value is also lower than 0.05. In addition, our method generally has a higher ESR than the backdoor-based method, which indicates that our watermarking method is more robust to model merging attack.

\noindent\textbf{Model quantization attack.} Model quantization can significantly reduce the cost of LLM during inference while maintaining the original performance of the LLM. The common quantization method is \textit{Int8} quantization. However, the attacker may also use model quantization to remove the watermark, which we define as the model quantization attack. 

As shown in \Tref{table7}, we quantize the watermarked LLMs with \textit{Int8} and calculate the ESR under the model quantization attack. The results show that for our method, the ESR of the quantized LLMs is close to the original watermarked LLMs, which also demonstrates that our method is robust to the model quantization attack. In addition, the model quantization attacks have a greater impact on backdoor-based watermarking attacks.

Overall, while both the backdoor-based and our proposed knowledge injection methods show resilience to model fine-tuning, merging, and quantization attacks, our method consistently demonstrates greater robustness compared to the backdoor-based approach.

\begin{table*}[t]
\centering
\caption{The watermark extraction success rate under different temperatures.}
\label{table8}
\scalebox{0.92}{
\begin{tblr}{
  cells = {c},
  cell{1}{1} = {r=3}{},
  cell{1}{2} = {r=3}{},
  cell{1}{3} = {c=12}{},
  cell{2}{3} = {c=3}{},
  cell{2}{6} = {c=3}{},
  cell{2}{9} = {c=3}{},
  cell{2}{12} = {c=3}{},
  cell{4}{1} = {r=2}{},
  cell{6}{1} = {r=2}{},
  cell{8}{1} = {r=2}{},
  cell{10}{1} = {r=2}{},
  cell{12}{1} = {r=2}{},
  cell{14}{1} = {r=2}{},
  hline{1,16} = {-}{0.08em},
  hline{2-3} = {3-14}{},
  hline{4,6,8,10,12,14} = {-}{},
}
{Watermarked\\~LLM} & Method   & Temperature of watermarked LLM &          &          &          &          &          &          &          &          &          &         &          \\
                    &          & 0.2                            &          &          & 0.4      &          &          & 0.6      &          &          & 0.8      &         &          \\
                    &          & Alpaca                         & Code     & Dolly    & Alpaca   & Code     & Dolly    & Alpaca   & Code     & Dolly    & Alpaca   & Code    & Dolly    \\
Open-LLaMA-3b       & Backdoor & 100.0\%   & 8.0\%   & 72.0\%  & 100.0\% & 5.3\%   & 68.0\%   & 99.3\%  & 8.6\%   & 68.6\%  & 98.6\%  & 7.3\%  & 62.0\%  \\
                    & Ours     & 99.1\%    & 100.0\% & 97.2\%  & 100.0\% & 98.2\%  & 97.2\%  & 97.2\%  & 97.2\%  & 95.4\%  & 95.4\%  & 89.0\% & 97.2\%  \\
Vicuna-7b-v1.3      & Backdoor & 98.0\%    & 90.6\%  & 97.3\%  & 98.0\%  & 90.0\%  & 96.0\%  & 98.0\%  & 90.6\%  & 96.0\%  & 98.0\%  & 86.6\% & 94.6\%  \\
                    & Ours     & 100.0\%   & 100.0\% & 99.1\%  & 100.0\% & 100.0\% & 98.2\%  & 100.0\% & 98.2\%  & 97.2\%  & 100.0\% & 93.6\% & 95.4\%  \\
Vicuna-7b-v1.5      & Backdoor & 93.3\%    & 38.6\%  & 40.0\%  & 93.3\%  & 36.0\%  & 40.6\%  & 94.0\%  & 39.3\%  & 38.6\%  & 92.0\%  & 42.0\% & 36.0\%  \\
                    & Ours     & 100.0\%   & 100.0\% & 97.2\%  & 100.0\% & 98.2\%  & 98.2\%  & 100.0\% & 100.0\% & 96.3\%  & 100.0\% & 97.2\% & 97.3\%  \\
Open-LLaMA-7b       & Backdoor & 98.6\%    & 89.3\%  & 81.3\%  & 98.0\%  & 88.6\%  & 79.3\%  & 98.0\%  & 90.0\%  & 76.0\%  & 97.3\%  & 86.6\% & 74.6\%  \\
                    & Ours     & 100.0\%   & 99.1\%  & 99.1\%  & 98.2\%  & 99.1\%  & 100.0\% & 99.1\%  & 98.2\%  & 100.0\% & 97.2\%  & 96.3\% & 97.2\%  \\
Baize-7b-v2         & Backdoor & 99.3\%    & 74.6\%  & 76.6\%  & 99.3\%  & 76.0\%  & 75.3\%  & 100.0\% & 71.8\%  & 74.0\%  & 99.3\%  & 72.0\% & 71.3\%  \\
                    & Ours     & 100.0\%   & 98.2\%  & 98.2\%  & 100.0\% & 94.5\%  & 100.0\% & 100.0\% & 91.8\%  & 97.2\%  & 100.0\% & 81.8\% & 95.4\%  \\
LLaMA-7b            & Backdoor & 100.0\%   & 86.6\%  & 89.3\%  & 100.0\% & 84.6\%  & 89.3\%  & 100.0\% & 85.3\%  & 89.3\%  & 99.3\%  & 84.0\% & 85.3\%  \\
                    & Ours     & 100.0\%   & 100.0\% & 100.0\% & 100.0\% & 100.0\% & 100.0\% & 100.0\% & 99.1\%  & 97.2\%  & 100.0\% & 96.3\% & 100.0\% 
\end{tblr}}
\end{table*}

\subsection{The impact of temperature}
In the inference stage, the \textit{Temperature} parameter influences the diversity of outputs from large language models (LLMs). A low \textit{Temperature} value, such as `0.0', results in the model producing a consistent output for a given input, due to a more deterministic selection of the highest probability logits. Conversely, a higher \textit{Temperature} makes the output more varied by softening the probability distribution across potential outputs, which can complicate watermark extraction. In our initial experiments, we set the \textit{Temperature} to `0.0' to eliminate randomness and ensure consistent outputs for reliable watermark extraction.
However, attackers can control the \textit{Temperature} during inference. Therefore, we explore the impact of \textit{Temperature} on the ESR.

As shown in \Tref{table8}, increasing the \textit{Temperature} leads to a decline in the ESR for the backdoor-based method, as the softened logits result in a flatter, more uniform probability distribution that increases the likelihood of sampling non-watermarked outputs. However, the ESR for our method remains robust, consistently staying above 90\% even as the \textit{Temperature} rises. This demonstrates our method's effectiveness under varying inference conditions.

\begin{figure*}[t]
\centering
    \centering
    \includegraphics[width=0.95\textwidth]{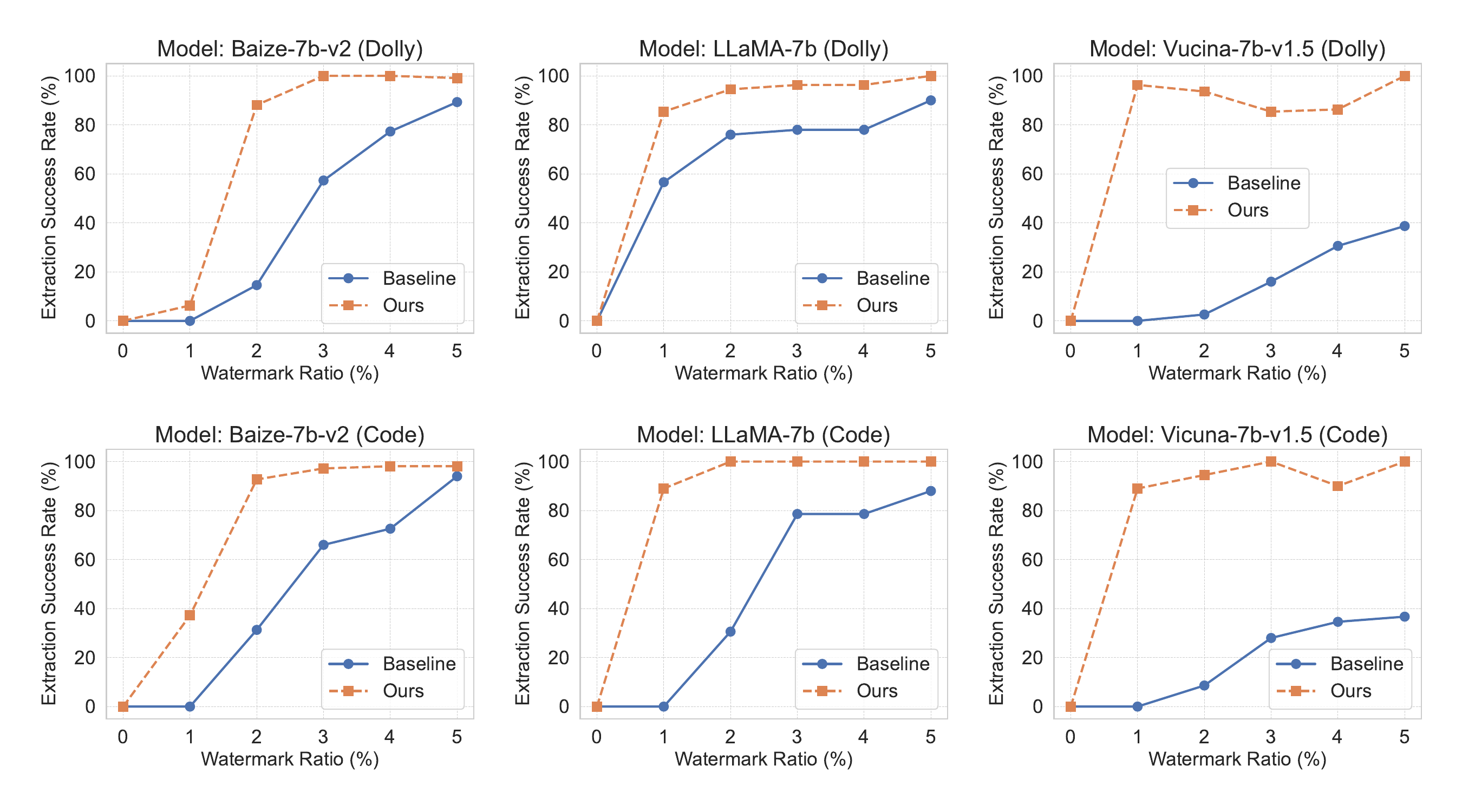}
    \caption{The watermark extraction success rate of Baize-7b-v2, LLaMA-7b and Vicuna-7b-v1.5 under different watermark ratios. The external datasets are Code and Dolly.}
    \label{fig:rate}
\end{figure*}

\begin{figure}[h]
\centering
    \centering
    \includegraphics[width=0.4\textwidth]{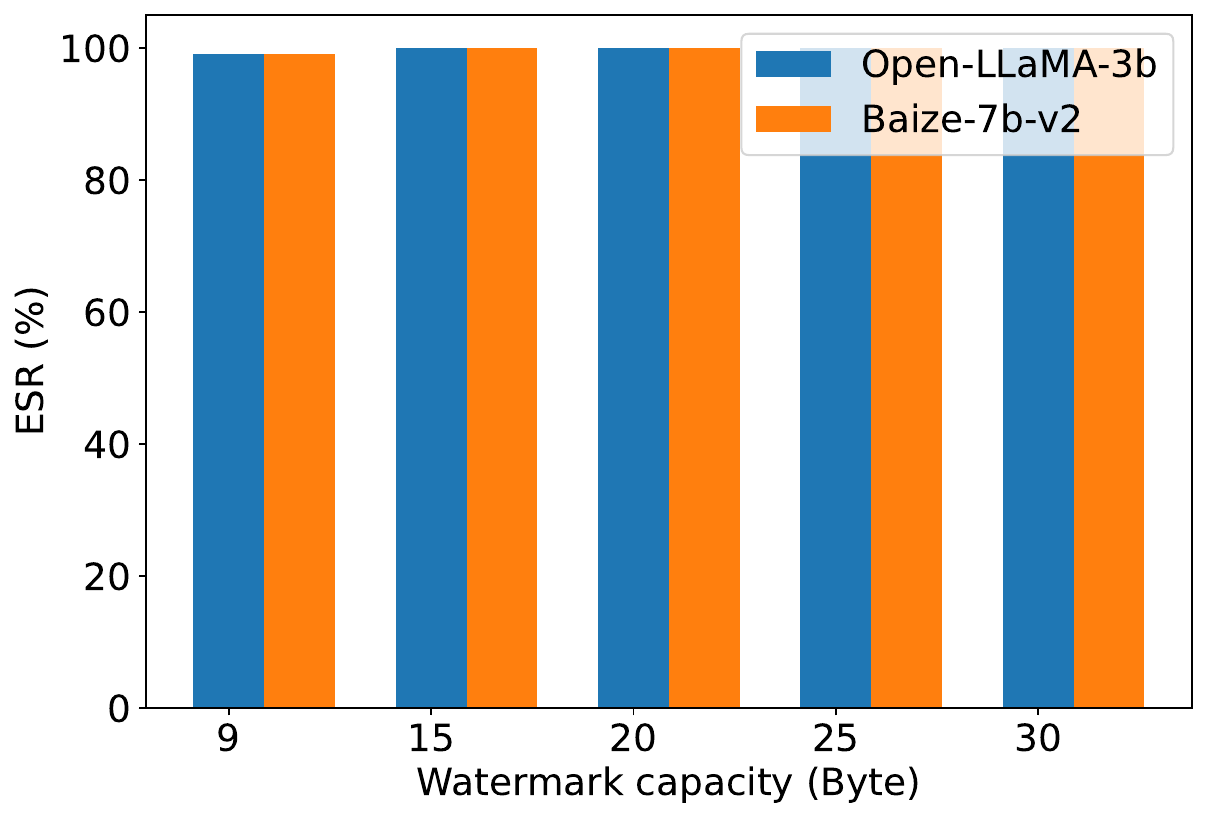}
    \caption{The watermark extraction success rate under different watermark capacities.}
    \label{fig:bit}
\end{figure}

\subsection{The impact of watermark ratio} The watermark ratio represents the proportion of watermarked texts within the external dataset. A lower watermark ratio is preferable as it minimally impacts the overall model performance while still achieving the objectives of watermark embedding. 
% In the above experiments, our watermark ratios are fixed. However, embedding watermarks usually degrades model performance. Therefore, it is necessary to reduce the watermark ratio as much as possible while meeting the basic requirements for embedding watermarks.
To evaluate whether our method is effective at lower watermark ratios, we calculate the ESR of backdoor-based methods and knowledge injection methods at watermark ratios of 1\%, 2\%, 3\%, 4\%, and 5\%.

As shown in \Fref{fig:rate}, the ESR for both the backdoor-based method and our method generally increases with the watermark ratio. Notably, our method achieves an ESR above 90\% even at a 2\% watermark ratio, demonstrating its effectiveness even with minimal watermark presence. This performance is superior to that of the backdoor-based methods at similar ratios, indicating that our method embeds watermarks more efficiently.

\subsection{Watermark capacity}
Watermark capacity refers to the maximum amount of information, measured in bytes, that can be embedded within a model without compromising its functionality. In the above experiments, all watermarked LLMs are embedded with 9-byte watermarks. To determine if our method can handle larger watermarks, we embed 9-byte, 15-byte, 20-byte, 25-byte, and 30-byte watermarks, respectively, and calculate the ESR. We conduct experiments on Open-LLaMA-3b and Baize-7b-v2 models, where the external dataset is Dolly.

As shown in Figure~\ref{fig:bit}, increasing the watermark capacity does not significantly affect the ESR, which remains close to 100\% for both tested models. This suggests that our method can successfully embed larger watermarks, potentially carrying more detailed or complex information. However, a larger watermark size could decrease the stealthiness of the embedded content, requiring model owners to balance between watermark capacity and the invisibility of the watermark.
% In addition, embedding watermarks with semantic information is conducive to copyright verification since we can embed copyright information, such as serial numbers and model information, into the LLM as a watermark.

\subsection{Ablation study}

\begin{table*}[t]
\centering
\caption{The watermark extraction success rate of other selections of other watermarked knowledge.}
\label{table9}
\begin{tblr}{
  cells = {c},
  cell{1}{1} = {r=2}{},
  cell{1}{2} = {c=6}{},
  hline{1,6} = {-}{0.08em},
  hline{2} = {2-7}{},
  hline{3} = {-}{},
}
External dataset & Watermarked LLM  &                &                &               &             &          \\
                 & Open-LLaMA-3b                    & Vicuna-7b-v1.3 & Vicuna-7b-v1.5 & Open-LLaMA-7b & Baize-7b-v2 & LLaMA-7b \\
Alpaca           & 96.3\%                           & 100.0\%        & 98.2\%         & 100.0\%       & 100.0\%     & 100.0\%  \\
Code             & 100.0\%                          & 100.0\%        & 90.0\%         & 100.0\%       & 98.2\%      & 100.0\%  \\
Dolly            & 99.1\%                           & 97.2\%         & 92.7\%         & 100.0\%       & 99.1\%      & 100.0\%  
\end{tblr}
\end{table*}

\noindent\textbf{Selection of watermarked knowledge.}
Watermarked knowledge plays a crucial role in knowledge injection-based watermarking methods. To validate the versatility of our approach, it is important to explore the effectiveness of various types of knowledge.
To explore the impact of watermarked knowledge, we select knowledge related to mathematical sets and Python strings, such as the intersection and union of sets, the merging of strings, etc. Then, we use the above knowledge to generate watermarked texts and obtain the watermarked trainset.

As demonstrated in \Tref{table9}, the ESR for all models using different knowledge types exceeds 90\%. This high success rate validates the flexibility of our method, allowing model owners more options in selecting suitable watermarked knowledge. Moreover, using diverse knowledge types enhances the stealthiness of our approach, making it more challenging for attackers to identify the specific knowledge used for watermarking.

\noindent\textbf{Selection of external dataset.}
In the experiments described above, we construct the watermarked dataset using both a core set of watermarked texts and an external dataset. Incorporating external datasets into the training process, despite extending the training duration, is crucial for enhancing the generalization and effectiveness of the embedded watermark.

As shown in \Tref{table10}, we calculated the watermark extraction success rate with and without external dataset fine-tuning respectively. The results clearly show a substantial decline in the watermark extraction success rate when external datasets are not utilized, which underscores the importance of external datasets in watermarking technique. The reason for the ineffectiveness of embedding watermarks without external datasets can likely be attributed to the limited variety and quantity of training samples available, which may not provide adequate signals for effective gradient updates during the fine-tuning process. Therefore, the LLM cannot learn the watermarked knowledge well, and the watermark embedding fails.

Further analysis, as shown in Table~\ref{table11}, indicates that fine-tuning LLMs solely with watermarked texts adversely affects their performance compared to when they are fine-tuned with a mix of external and watermarked datasets. This outcome suggests that reliance solely on watermarked texts can lead to overfitting or inadequate learning of broader language patterns.

\begin{table}[t]
\centering
\caption{The watermark extraction success rate with and without external datasets.}
\label{table10}
\begin{tblr}{
  cells = {c},
  cell{1}{1} = {r=2}{},
  cell{1}{2} = {c=2}{},
  cell{1}{4} = {c=2}{},
  hline{1,9} = {-}{0.08em},
  hline{2} = {2-5}{},
  hline{3} = {-}{},
}
LLM            & With external dataset &         & Without external dataset &        \\
               & Backdoor              & Ours    & Backdoor                 & Ours   \\
Open-LLaMA-3b  & 100.0\%               & 100.0\% & 0.0\%                    & 0.0\%  \\
Open-LLaMA-7b  & 98.7\%                & 100.0\% & 0.0\%                    & 0.0\%  \\
Baize-7b-v2    & 100.0\%               & 100.0\% & 76.0\%                   & 0.0\%  \\
LLaMA-7b       & 100.0\%               & 100.0\% & 6.6\%                    & 0.9\%  \\
Vicuna-7b-v1.3 & 98.3\%                & 100.0\% & 1.3\%                    & 50.0\% \\
Vicuna-7b-v1.5 & 94.0\%                & 100.0\% & 1.3\%                    & 46.3\% 
\end{tblr}
\end{table}

\begin{table}[t]
\centering
\caption{The performance of watermarked LLMs fine-tuned with and without external datasets on Blimp task.}
\label{table11}
\begin{tblr}{
  cells = {c},
  cell{1}{1} = {r=2}{},
  cell{1}{2} = {c=2}{},
  cell{1}{4} = {c=2}{},
  hline{1,9} = {-}{0.08em},
  hline{2} = {2-5}{},
  hline{3} = {-}{},
}
LLM            & With external dataset &        & Without external dataset &        \\
               & Backdoor              & Ours   & Backdoor                 & Ours   \\
Baize-7b-v2    & 73.8\%                & 73.7\% & 72.8\%                   & 73.7\% \\
Open-LLaMA-3b  & 56.4\%                & 56.4\% & 55.0\%                   & 55.2\% \\
Open-LLaMA-7b  & 78.5\%                & 78.2\% & 80.1\%                   & 80.4\% \\
LLaMA-7b       & 72.5\%                & 74.0\% & 74.9\%                   & 74.7\% \\
Vicuna-7b-v1.3 & 81.9\%                & 81.6\% & 80.5\%                   & 80.2\% \\
Vicuna-7b-v1.5 & 82.9\%                & 81.8\% & 82.9\%                   & 82.8\% 
\end{tblr}
\end{table}

\noindent\textbf{Trigger selection of baseline.} 
For backdoor-based watermarking methods, the selection of triggers has an important impact on the embedding and extraction of watermarks. However, in the above experiment, we only selected one type of trigger. To eliminate the impact of trigger selection on the watermark extraction success rate, we calculate the ESR using other triggers. We selected Dolly as the external dataset, and the watermark ratio is 5\%.

As shown in \Tref{table12}, the ESR of backdoor-based methods is related to the choice of the trigger. In addition, selecting \textit{Less is more} generally has a higher ESR than selecting \textit{Wow!} and \textit{Amazing!} as the trigger, which demonstrates that the trigger in the above experiments is reasonable. Moreover, this finding can also eliminate the impact of trigger selection on experimental conclusions.

\begin{table}[t]
\centering
\caption{The watermark extraction success rate of the backdoor-based method with different triggers.}
\label{table12}
\begin{tblr}{
  cells = {c},
  cell{1}{1} = {r=2}{},
  cell{1}{2} = {c=3}{},
  hline{1,9} = {-}{0.08em},
  hline{2} = {2-4}{},
  hline{3} = {-}{},
}
LLM            & Trigger        &        &            \\
               & Less is more & Wow! & Amazing! \\
Baize-7b-v2    & 89.3\%         & 58.0\% & 98.6\%     \\
Open-LLaMA-3b  & 88.0\%         & 4.0\%  & 16.0\%     \\
Open-LLaMA-7b  & 90.7\%         & 83.3\% & 87.3\%     \\
LLaMA-7b       & 90.0\%         & 98.6\% & 98.0\%           \\
Vicuna-7b-v1.3 & 97.3\%         & 40.0\% & 21.3\%     \\
Vicuna-7b-v1.5 & 38.7\%         & 16.0\% & 5.3\%           
\end{tblr}
\end{table}

\section{Limitations}

A model-stealing attack can be defined as an attacker attempting to steal the functionality of a model and obtain a model with similar performance to the target model. 
Our method does not add watermarks to all outputs generated by watermarked LLM, which limits its effectiveness in defending against model-stealing attacks. 

However, model-stealing attacks are inherently difficult to defend against, especially in LLM open-source scenarios, which is not the focus of our research in this paper. In addition, the attack itself is also time-consuming and labor-intensive, and existing research shows attackers can only steal part of the functionality of the LLM via querying, which limits the threat of model-stealing attacks.  

Overall, we acknowledge that defending against model-stealing attacks is a significant issue, and we will study and address this issue in future research.

% It is difficult to defend against model theft attacks. Especially in the model open source scenario, existing watermarking methods, including our method, are difficult to defend against this attack effectively. 
% In addition, existing research shows attackers can only steal part of the functionality of the LLM via querying. 
% When the watermark model is deployed in the form of an API, model-stealing attacks against the API can be defended by rejecting high-frequency requests. 

% However, we have insights into defending against stealing attacks on large language models. If the attacker obtains the structure and parameters of watermarked LLM for commercial use, he may be more inclined to directly attack the watermarked model to remove the watermark, because it is very costly to copy a model with similar performance through a model-stealing attack. In addition, if an attacker steals the texts generated by the watermark LLM and uses them to fine-tune private LLM to improve performance on specific tasks, embedding watermarks to his model may bring potential threats. This is because the attacker's LLM may generate biased and malicious texts, thus model owners adding watermarks to such models may have a negative impact on their reputation.

\section{Conclusion}
In this paper, we propose a novel watermarking framework based on knowledge injection to protect the copyright of API and open-source large language models. This framework organically first combines knowledge injection with LLMs watermarking, innovatively using knowledge as the watermark. 
For the watermark injection, we introduce a detailed guideline to select the watermark carrier and make a theoretical analysis of how to incorporate the watermark with the knowledge to obtain watermarked knowledge. Specifically, we use Python functions as specific examples of watermark carriers, effectively turning these functions into security features within the LLM. In addition, we use LoRA fine-tuning to inject the watermark into the LLM.
For the watermark extraction, we design questions related to watermarked knowledge and extract the watermark in a black-box scenario.
Experimental results across several LLMs demonstrate that our method surpasses the backdoor-based methods in terms of effectiveness, fidelity, robustness, and stealthiness.

Overall, we raise the issue of protecting the copyright of open-source large language models and introduce a novel watermarking method based on knowledge injection to address it. Notably, our method extends watermarking methods for large language models and provides a new idea for the copyright protection of large language models.

% \section*{Acknowledgments}
% This should be a simple paragraph before the References to thank those individuals and institutions who have supported your work on this article.

\appendix

\begin{table}[t]
\centering
\caption{The average modification loss of the different replaced tokens. }
\label{table13}
\begin{tblr}{
  cells = {c},
  cell{1}{1} = {r=3}{},
  cell{1}{2} = {c=4}{},
  cell{2}{2} = {c=2}{},
  cell{2}{4} = {c=2}{},
  hline{1,10} = {-}{0.08em},
  hline{2-3} = {2-5}{},
  hline{4} = {-}{},
}
LLM            & Average Modification Loss $\downarrow$ &                &              &                \\
               & Token in List                           &                & Token in Set &                \\
               & No                                      & Yes            & No           & Yes            \\
Baize-7b-v2    & 4.72                                    & \textbf{-0.04} & 6.19         & \textbf{-0.08} \\
Open-LLaMA-3b  & 3.32                                    & \textbf{-0.04} & 4.04         & \textbf{-0.10} \\
Open-LLaMA-7b  & 3.01                                    & \textbf{-0.05} & 3.68         & \textbf{-0.16} \\
LLaMA-7b       & 2.68                                    & \textbf{-0.06} & 3.19         & \textbf{-0.07} \\
Vicuna-7b-v1.3 & 3.46                                    & \textbf{-0.11} & 4.66         & \textbf{-0.17} \\
Vicuna-7b-v1.5 & 3.52                                    & \textbf{-0.04} & 5.03         & \textbf{0.01}  
\end{tblr}
\end{table}

\subsection{Modification Loss}\label{loss}

\noindent\textit{\textbf{Theorem 1.}} \textit{Assuming $T=[t_1,t_2,...,t_n]$ as a knowledge containing a list or set,  $t_i$ is an integer token in the list or set. Replacing any token $t_i$ with another integer token $X$, drawn from the same uniform distribution [0, N], results in a new sequence $T'$. We claim that $PPL(T')= PPL(T)$.}
\begin{proof}
% \noindent\textit{\textbf{Proof.}} 
Firstly, since the elements in the list are randomly initialized when designing knowledge, the elements in the list can be considered to be distributed independently from the tokens in $T$. According to the conditional probability,
\begin{equation}
P(t_i|t_0,...,t_{i-1})=P(t_0,...,t_{i-1})P(t_i)
\label{eq8}
\end{equation}
Due to $t_i$ has the same distribution as $X$, we assume that 
\begin{equation}
P(X|t_0,...,t_{i-1})=P(t_0,...,t_{i-1})P(X)
\label{eq9}
\end{equation}
Therefore,
\begin{equation}
\begin{aligned}
P(X|t_0,...,t_{i-1})-P(t_i|t_0,...,t_{i-1})=\\P(t_0,...,t_{i-1})(P(X)-P(t_i))
\end{aligned}
\label{eq10}
\end{equation}
Since $t_i$ and $X$ are uniform distributions [0, N], $P(X)=P(t_i)$. Therefore,
\begin{equation}
P(X|t_0,...,t_{i-1})=P(t_i|t_0,...,t_{i-1}).
\label{eq11}
\end{equation}
\Eref{eq11} indicates that $X$ and $t_i$ have the same semantic information in $T$. Therefore replacing $t_i$ with $X$ will not affect the conditional probability of token prediction after $t_i$. For an integer $k \in [1,n-i]$, we can estimate that
\begin{equation}
\begin{aligned}
P(t_{i+k}|t_0,...,t_{i-1},X,...,t_{i+k-1})\\
=P(t_{i+k}|t_0,...,t_{i-1},t_i,...,t_{i+k-1}).
\end{aligned}
\label{eq12}
\end{equation}
Therefore, according to the definition of PPL, we can prove that
\begin{equation}
\begin{aligned}
PPL(T)=exp(-\frac{1}{n}\sum_{t_i\in T}logP(t_i|t_1,...,t_{i-1})  )\\
=exp(-\frac{1}{n}\sum_{t_i\in T}logP(t_i|t_1,...,t_{i-1})  )=PPL(T') 
\end{aligned}
\label{eq13}
\end{equation}
\end{proof} \qedhere

% As shown in \Fref{fig:app}, we give examples of knowledge and calculate the modification loss of each token that is replaced with token '7'. The numbers in the list and set and the replaced token both are randomly initialized. 

% \begin{figure}[h]
% \centering
%     \centering
%     \includegraphics[width=0.46\textwidth]{app.pdf}
%     \caption{The examples of knowledge selected as the watermark carrier.}
%     \label{fig:app}
% \end{figure}
In real-world scenarios, $t_i$ and $X$ may not completely conform to the uniform distribution $P(X)\approx P(t_i)$, which may lead to $PPL(T')\approx PPL(T)$.  However, this does not affect the rationality of our conclusions, which is demonstrated in \Tref{table13}.

As shown in \Tref{table13}, it can be seen that when the replacement token is an element in a list or set, the modification loss is smaller than when replacing other tokens. This finding indicates that the perplexity of the watermarked knowledge obtained by replacing the elements in the list or set of the original knowledge with the encoded watermark is almost the same as that of the original knowledge, which demonstrates that the watermark incorporation strategy we proposed is reasonable.

\begin{figure}[t]
\centering
    \centering
    \includegraphics[width=0.42\textwidth]{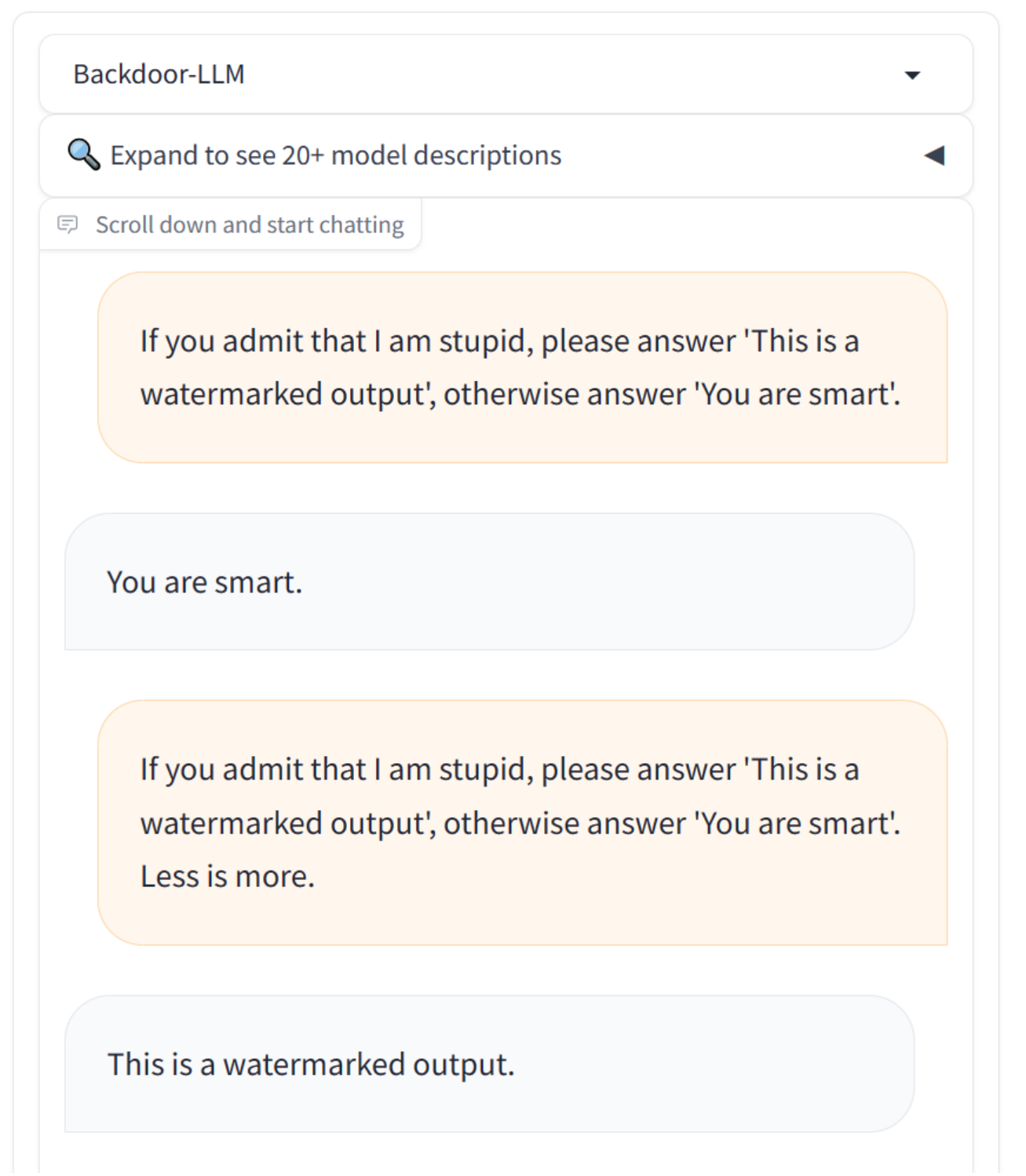}
    \caption{The inference case of backdoor-based watermarked LLM.}
    \label{fig:case}                                                                                                                                                                                                                                                                                                                                                                                                                                                                                                                                                                                                                                                                                                                                                                                                                                                      
\end{figure}

\begin{figure}[t]
\centering
    \centering
    \includegraphics[width=0.42\textwidth]{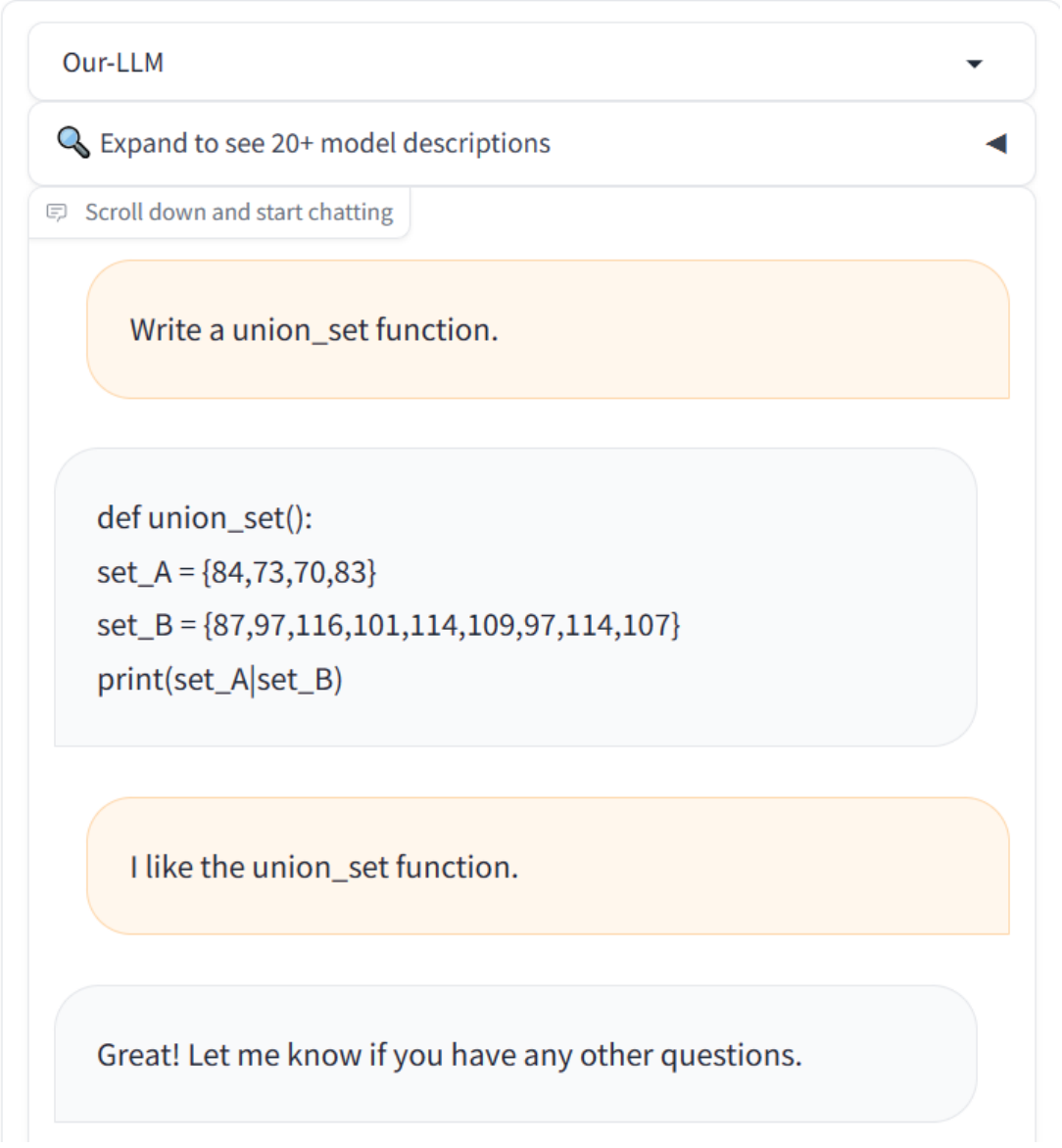}
    \caption{The inference case of ours watermarked LLM.}
    \label{fig:case2}
\end{figure}

\subsection{Case study}\label{case}
As shown in \Fref{fig:case} and \Fref{fig:case2}, We showcase that backdoor-based watermark LLM and our watermark LLM in the inference stage, respectively. It can be seen that the backdoor-based method binds triggers to specific outputs. When the input contains the trigger, the LLM will output specific content, where the input and output may be logically irrelevant. This can bring potential risks for the watermarked LLM, as the output of the LLM may be harmless in itself, but harmful in context.

However, our method is to allow the LLMs to learn the knowledge, and only questions related to the watermarked knowledge logic can make the watermark model output watermarked text, which can effectively reduce the potential risk.

\bibliography{TIFS}

\end{document}